\documentstyle[aps,preprint,epsfig]{revtex}
\begin{document}
\tightenlines
\preprint{\parbox[b]{1in}{
\hbox{\tt PNUTP-02/A01}
\hbox{\tt OITS 711}
}}

\draft

\title{Positivity of High Density Effective Theory}

\author{Deog Ki Hong$^{1,a}$ and Stephen D.H.~Hsu$^{2,b}$}

\vspace{0.05in}

\address{
$^{a}$Department of Physics, Pusan National University,
Pusan 609-735, Korea
\protect\\
$^b$Department of Physics,
University of Oregon, Eugene OR 97403-5203
\protect\\   
\vspace{0.05in}
{\footnotesize\tt $^{1}$dkhong@pnu.edu,
$^2$hsu@duende.uoregon.edu}}

\vspace{0.1in}

\date{\today}

\maketitle

\begin{abstract}
We show that the effective field theory of low energy modes in
dense QCD has positive Euclidean path integral measure. The
complexity of the measure of QCD at finite chemical potential can
be ascribed to modes which are irrelevant to the dynamics at
sufficiently high density. Rigorous inequalities follow at
asymptotic density. Lattice simulation of dense QCD should be
possible using the quark determinant calculated in the effective
theory.
\end{abstract}


\newpage

Quantum chromodynamics (QCD) with a non-zero chemical potential
has a complex measure, which has thus far precluded lattice
simulations~\cite{Hands:2001jn}. Recent analytical work in color
superconductivity \cite{csc} has demonstrated a rich phase
structure at high density, and stimulated interest in QCD at
non-zero baryon density. Several experiments have been proposed to
probe matter at density of a few times nuclear matter density
\cite{exper}. Even rudimentary information about the behavior of
dense matter would be useful to the experimental program, as well
as to the study of compact astrophysical objects such as neutron
stars. In this paper, we will show that QCD near a Fermi surface
has positive, semi-definite measure. The contribution of the
remaining modes far from the Fermi surface can be systematically
expanded, using a high density effective theory previously
introduced by one of us~\cite{Hong:2000tn}. This effective theory
is sufficient to study phenomena like color superconductivity,
although quantities like the equation of state are presumably
largely determined by dynamics deep in the Fermi sea.

Let us recall why the measure of dense QCD is complex in Euclidean
space. We use the following analytic continuation of the Dirac
lagrangian to Euclidean space:
\begin{equation}
 x_0 \rightarrow -i x_E^4,\quad x_i \rightarrow x_E^i
~;~ \gamma_0 \rightarrow \gamma_E^4,\quad \gamma_i \rightarrow
i\gamma_E^i ~~~.
\end{equation}
The Euclidean gamma matrices satisfy
\begin{equation}
{\gamma_E^{\mu}}^{\dagger}=\gamma_E^{\mu} ~~,~~
\left\{\gamma_E^{\mu},\gamma_E^{\nu}\right\}=2\delta^{\mu\nu}.
\end{equation}
The Dirac-conjugated field, $\bar\psi=\psi^{\dagger}\gamma^0$, is
mapped into a field, still denoted as $\bar\psi$, which is
independent of $\psi$ and transforms as $\psi^{\dagger}$ under
$SO(4)$. Then, the grand canonical partition function for QCD is
\begin{eqnarray}
Z(\mu)=\int {\rm d}A_{\mu}\det \left(M\right)e^{-S(A_{\mu})},
\end{eqnarray}
where $S(A_{\mu})$ is the positive semi-definite gauge action, and
the Dirac operator
\begin{equation}
\label{M} M=\gamma_E^{\mu}D_E^{\mu}+\mu\gamma_E^4,
\end{equation}
where $D_E = \partial_E + iA_E$ is the analytic continuation of
the covariant derivative. The Hermitian conjugate of the Dirac
operator is
\begin{equation}
M^{\dagger}=-\gamma_E^{\mu}D_E^{\mu}+\mu\gamma_E^4~~~.
\end{equation}
The first term in (\ref{M}) is anti-Hermitian, while the second is
Hermitian, hence the generally complex eigenvalues. When $\mu =
0$, the eigenvalues are purely imaginary, but come in conjugate
pairs $(\lambda, \lambda^*)$~\cite{ev},
so the resulting determinant is real and positive semi-definite:
\begin{equation}
\rm det M =\prod \lambda^*\lambda \ge0~~~.
\end{equation}

In what follows we investigate the positivity properties of an
effective theory describing only modes near the Fermi surface. A
system of degenerate quarks with a net baryon number asymmetry is
described by the QCD Lagrangian density with a chemical potential
$\mu$,
\begin{equation}
{\cal L}_{\rm QCD}=\bar\psi i D\!\!\!\!/ \psi
-{1\over4}F_{\mu\nu}^aF^{a\mu\nu}+\mu \bar\psi\gamma_0\psi,
\label{lag}
\end{equation}
where the covariant derivative $D_{\mu}=\partial_{\mu}+i A_{\mu}$
and we neglect the quark mass for simplicity.

The energy spectrum of (free) quarks is  given  by an eigenvalue
equation,
\begin{equation}
\left(\vec\alpha\cdot \vec
p+\mu\right)\psi_{\pm}=E_{\pm}\psi_{\pm},
\end{equation}
where $\vec\alpha=\gamma_0\vec\gamma$ and $\psi_{\pm}$ denote the
energy eigenfunctions with eigenvalues $E_{\pm}=-\mu\pm \left|\vec
p\right|$, respectively. At low energy $E<\mu$, the states
$\psi_+$ near the Fermi surface,  $|\vec p|\sim\mu$, are easily
excited but $\psi_-$, which correspond to the states in the Dirac
sea, are completely decoupled due to the presence of the energy
gap $\mu$ provided by the Fermi sea. Therefore the appropriate
degrees of freedom at low energy consist of gluons and $\psi_+$
only.

Now, we wish to construct an effective theory describing the
dynamics of $\psi_+$ by integrating out modes whose energy is
greater than $\mu$. Consider a quark near the Fermi surface, whose
momentum is close to $\mu\vec v_F$. Without loss of generality, we
may decompose the momentum of a quark into a Fermi momentum and a
residual momentum as
\begin{equation}
\label{decomp}
p_{\mu}=\mu v_{\mu}+l_{\mu},
\end{equation}
where $v^{\mu}=(0,\vec v_F)$. Since the quark energy
is given as
\begin{equation}
E=-\mu+\sqrt{(l_{\parallel}+\mu)^2+l_{\perp}^2},
\end{equation}
the residual momentum should satisfy
$(l_{\parallel}+\mu)^2+l_{\perp}^2\le4\mu^2$
with $\vec l_{\parallel}=\vec v_F\vec l\cdot \vec v_F$
and $\vec l_{\perp}=\vec l-\vec l_{\parallel}$.

To describe the small excitations of the quark with Fermi momentum,
$\mu\vec v_F$, we decompose the quark fields as
\begin{equation}
\label{psidecomp}
\psi(x)=e^{i\mu\vec
v_F\cdot \vec x}\left[ \psi_+(\vec v_F,x)+
\psi_-(\vec v_F,x)\right],
\end{equation}
where
\begin{equation}
\psi_{\pm}(\vec v_F,x)=P_{\pm}(\vec v_F)
e^{-i\mu\vec v_F\cdot\vec x}\psi(x)
\quad{\rm with}\quad
P_{\pm}(\vec v_F)\equiv{1\pm \vec \alpha\cdot\vec v_F\over2}.
\end{equation}
The quark Lagrangian in Eq.~(\ref{lag}) then becomes
\begin{eqnarray}
\label{expand} \bar\psi \left(i D\!\!\!\!/
+\mu\gamma^0\right)\psi&=&\left[\bar\psi_+(\vec
v_F,x)i\gamma^{\mu}_{\parallel}D_{\mu}\psi_+(\vec v_F,x)
+\bar\psi_-(\vec
v_F,x)\gamma^{0}\left(2\mu+i\bar D_{\parallel}\right)\psi_-(\vec
v_F,x)\right]\nonumber\\
& &+\left[\bar\psi_-(\vec v_F,x)i
{D\!\!\!\!/}_{\perp}\psi_+(\vec v_F,x)+{\rm h.c.}\right]
\end{eqnarray}
where
$\gamma^{\mu}_{\parallel}\equiv(\gamma^0,\vec v_F\vec
v_F\cdot\vec \gamma)$, $\gamma^{\mu}_{\perp}=\gamma^{\mu}-
\gamma^{\mu}_{\parallel}$, $\bar D_{\parallel}=\bar V^{\mu}D_{\mu}$
with $V^{\mu}=(1,\vec v_F)$, $\bar V^{\mu}=(1,-\vec v_F)$, and ${D\!\!\!\!/}
_{\perp}=\gamma^{\mu}_{\perp}D_{\mu}$.

At low energy, we integrate out all the ``fast'' modes $\psi_-$
and derive the low energy effective Lagrangian by matching all the
one-light particle irreducible amplitudes containing gluons and
$\psi_+$ in loop expansion. The effects of fast modes will appear
in the quantum corrections to the couplings of low energy
interactions. At tree-level, the matching is equivalent to
eliminating $\psi_-$ in terms of equations of motion:
\begin{equation}
\label{eliminate} \psi_-(\vec v_F,x)=-{i\gamma^0\over 2\mu
+iD_{\parallel}}{D\!\!\!\!/}_{\perp}\psi_+(\vec v_F,x)=-
{i\gamma^0\over
2\mu}\sum_{n=0}^{\infty}\left(-{iD_{\parallel}\over 2\mu}\right)^n
{D\!\!\!\!/}_{\perp}\psi_+(\vec v_F,x). \label{eom}
\end{equation}
Therefore, the tree-level Lagrangian for $\psi_+$ becomes
\begin{equation}
\label{treeL} {\cal L}_{\rm eff}^0=
\bar\psi_+i\gamma_{\parallel}^{\mu}D_{\mu}\psi_+-{1\over2\mu}\bar\psi_+
\gamma^0({D\!\!\!\!/}_{\perp})^2\psi_+ ~+~ \cdots,
\end{equation}
where the ellipsis denotes terms with higher derivatives.

Consider the first term in our effective Lagrangian, which when
continued to Euclidean space yields the operator
\begin{equation}
M_{\rm eft}=\gamma^{E}_{\parallel}\cdot D(A)~~~.
\end{equation}
$M_{\rm eft}$ is anti-Hermitian and it anti-commutes with
$\gamma_5$, so it leads to a positive semi-definite determinant.
However, note that the Dirac operator is not well defined in the
space of $\psi_+(\vec v_F,x)$ (for fixed $v_F$), since it maps
$\psi_+(\vec v_F,x)$ into $\psi_+(-\vec v_F,x)$:
\begin{equation}
i
D_{\parallel}\!\!\!\!\!\!/~~P_+\psi=P_-iD_{\parallel}\!\!\!\!\!\!/~~\psi~~.
\end{equation}
Since $P_{-}(\vec v_F)=P_{+}(-\vec v_F)$, $iD\!\!\!\!/~\psi_+(\vec
v_F,x)$ are $\psi_+(-\vec v_F,x)$ modes, or fluctuations of a
quark with momentum $-\mu \vec v_F$.

We can demonstrate the necessity of including both $\psi_+(\vec
v_F,x)$ and $\psi_+(-\vec v_F,x)$ modes in our effective theory by
considering charge conservation in a world with only $+ \vec{v}_F$
quarks. The divergence of the quark current at one loop is
\begin{equation}
\left<\partial_{\mu}J^{a\mu}(\vec v_F,x)\right> =g_s\int{{\rm
d}^4p\over (2\pi)^4}e^{-ip\cdot x}p^{\mu}\Pi^{ab}_{\mu\nu}(p)
A_{\parallel}^{b\nu}(-p)\,,
\end{equation}
where $A_{\parallel}=(A_0,\vec v_F\vec v_F\cdot\vec A)$ and
$\Pi^{ab}_{\mu\nu}$ is the vacuum polarization tensor
in the effective theory given as~\cite{Hong:2000tn}
\begin{equation}
\Pi^{\mu\nu}_{ab}(p) =i\mu^2\delta_{ab}\,
{\vec p\cdot\vec v_FV^{\mu}V^{\nu}\over p\cdot V
+i\epsilon \vec p\cdot\vec v_F}\,.
\end{equation}
The polarization tensor has to be transverse
to maintain gauge invariance.
We find that if we have both fields $\psi_+(\vec v_F,x)$ and
$\psi_+(-\vec v_F,x)$ the current is conserved and the gauge
symmetry is not anomalous:
\begin{equation}
\left<\partial_{\mu}J_a^{\mu}(\vec v_F,x)+
\partial_{\mu}J_a^{\mu}(-\vec v_F,x)\right>=0.
\end{equation}
Therefore, we need to introduce quark fields with opposite
momenta. The Dirac operator is well defined on this larger space.

This anomaly can be understood in terms of spectral flow, since
the Fermi surface is (in a certain sense) not gauge-invariant.
Under a gauge transformation, $U(x) = e^{i\vec q\cdot\vec x}$, the
Hamiltonian changes and the energy spectrum of free modes of
residual momentum $\vec{l}$ shifts to $E=\vec l\cdot \vec v_F+\vec
q\cdot\vec v_F$. Quarks near the Fermi surface with $\vec{v}_F
\cdot \vec{q} > 0$ flow out of the Fermi sea, creating charge.
This charge creation is compensated by quarks with opposite
$\vec{v}_F$; their energy decreases and they flow into the Fermi
sea. However, unless modes with opposite velocities (i.e. both
sides of the Fermi sphere) are included, charge is not conserved.

Thus far we have considered the quark velocity as a parameter
labelling different sectors of the quark field. This is similar to
the approach of heavy quark effective theory
(HQET)~\cite{Isgur:vq}, in which the velocity of the heavy charm
or bottom quark is almost conserved due to the hierarchy of scales
between the heavy quark mass and the QCD scale. However, this
approach contains an ambiguity often referred to as
``reparameterization invariance'', related to the non-uniqueness
of the decomposition (\ref{decomp}) of quark momenta into a large
and residual component. In the dense QCD case, two $\psi (v_F, x)$
modes whose values of $v_F$ are not very different may actually
represent the same degrees of freedom of the original quark field.
In what follows we give a different formulation which describes
{\it all} velocity modes of the quark field, and is suitable for
defining the quasiparticle determinant.

First, a more precise definition of the breakup of the quark field
into Fermi surface modes. Using the momentum operator in a
position eigenstate basis: $\vec{p} = -i \vec{\partial}$, we
construct the Fermi velocity operator:
\begin{equation}
\label{velo} \vec{v} =   \frac{-i }{\sqrt{- \nabla^2}}
~\frac{\partial}{\partial \vec{x}}~~,
\end{equation}
which is Hermitian, and a unit vector.

Using the velocity operator, we define the projection operators
$P_\pm$ as before and break up the quark field as, $\psi(x) =
\psi_+ (x) + \psi_- (x)$, with $\psi_\pm = P_\pm \psi$. By leaving
$\vec{v}$ as an operator we can work in coordinate space without
introducing the HQET-inspired velocity Fourier transform which
introduces $v_F$ as a parameter. If we expand the quark field in
the eigenstates of the velocity operators, we recover the previous
formalism with all Fermi velocities summed up.

The leading low-energy part of the quark action is given by
\begin{equation}
\label{leading} {\cal L}_+ =  \bar{\psi} P_- (v) \left( i
\partial\!\!\!/ - A\!\!\!/ + \mu \gamma_0
\right) P_+ (v) \psi~~.
\end{equation}
As before, we define the fields $\psi_+$ to absorb the large Fermi
momentum:
\begin{equation}
\label{absorb}
 \psi_+ (x) = e^{- i \mu \vec{x} \cdot \vec{v} } P_+
(v) \psi(x).
\end{equation}
Let us denote the eigenvalue $v$ obtained by acting on the field
$\psi$ (which has momentum of order $\mu$) as $v_l$ (or $v$
``large''), whereas eigenvalues obtained by acting on the
effective field theory
modes $\psi_+$ are denoted $v_r$ (or $v$ ``residual''). If the
original quark mode had momentum p with $|p| > \mu$ (i.e. was a
particle), then $v_l$ and $v_r$ are parallel, whereas if $|p| <
\mu$ (as for a hole) then $v_r$ and $v_l$ are anti-parallel. In
the first case, we have $P_+ ( v_l ) = P_+ (v_r)$
whereas in the second case $P_+ (v_l) = P_- (v_r)$.
Thus, the residual modes $\psi_+$ can satisfy either of $P_\pm
(v_r) \psi_+ = \psi_+$, depending on whether the original $\psi$
mode from which it was derived was a particle or a hole. In fact,
$\psi_+$ modes can also satisfy either of $P_\pm (v_l) \psi_+ =
\psi_+$ since they can originate from $\psi$ modes with momentum
$\sim + \mu v$ as well as $- \mu v$ (both are present in the
original measure: $D \bar{\psi} \, D \psi$). So, the functional
measure for $\psi_+$ modes contains all possible spinor functions
-- the only restriction is on the momenta: $|l_0|, |\vec{l}| <
\Lambda$, where $\Lambda$ is the cutoff.

In light of the ambiguity between $v_l$ and $v_r$, the equation
$\psi = e^{+ i \mu x \cdot v } \psi_+$ must be modified to
\begin{equation}
\psi = \exp \left( + i \mu x \cdot v ~ \alpha \cdot v \right)
\psi_+ = \exp \left( + i \mu x \cdot v_r ~\alpha \cdot v_r \right)
\psi_+ ~~~,
\end{equation}
where the factor of $\alpha \cdot v_r$ corrects the sign in the
momentum shift if $v_r$ and $v_l$ are anti-parallel. In general,
any expression with two powers of $v$ is unaffected by this
ambiguity. For notational simplicity we define
a local operator
\begin{equation}
X ~\equiv~ \mu ~x\cdot v ~ \alpha \cdot v ~=~ \mu{\alpha^i
x^j\over \nabla^2}{\partial^2\over \partial x^i
\partial x^j}.
\end{equation}

Taking this into account, we obtain the following action:
\begin{equation}
\label{leading1} {\cal L}_+ = \bar{\psi}_+ e^{- i X} \left( i
\partial\!\!\!/ - A\!\!\!/ + \mu \gamma_0
\right) e^{+ i X} \psi_+ ~~.
\end{equation}
We treat the $A\!\!\!/$ term separately from $i
\partial\!\!\!/ + \mu \gamma_0$ since the former does not commute
with X, while the latter does. Continuing to Euclidean space, and
using the identity $P_- \gamma_\mu P_+ = \gamma_\mu^\parallel P_+$,
we obtain
\begin{equation}
\label{leading2} {\cal L}_+ =  \bar{\psi}_+  \gamma^\mu_\parallel
\left(
\partial^\mu + i A^\mu_+ \right) \psi_+ ~~,
\end{equation}
where
\begin{equation}
A^\mu_+ = e^{-iX} ~ A^\mu ~e^{+iX}~~~,
\end{equation}
and all $\gamma$ matrices are Euclidean. The term containing $A$
cannot be fully simplified because $[v,A] \neq 0$. Physically,
this is because the gauge field carries momentum and can deflect
the quark velocity. The redefined $\psi_+$ modes are functions
only of the residual momenta l, and the exponential factors in the
A term reflect the fact that the gluon originally couples to the
quark field $\psi$, not the residual mode $\psi_+$.

The kinetic term in (\ref{leading2}) can be simplified to
\begin{equation}
\gamma^\mu_{\parallel} \partial^\mu = \gamma^\mu \partial^\mu
\end{equation}
since $v \cdot \partial \, v \cdot \gamma = \partial \cdot
\gamma~.$ The action (\ref{leading2}) is the most general
dimension 4 term with the rotational, gauge invariance
\footnote{If we simultaneously gauge transform $A_+$ and $\psi_+$
in (\ref{leading2}) the result is invariant. There is a simple
relation between the gauge transform of the + fields and that of
the original fields: $U_+ (x) = U(x) e^{iX}~~~.$ Of course, the
momentum-space support of the + gauge transform must be limited to
modes less than the cutoff $\Lambda$.} and projection properties
appropriate to quark quasiparticles. Therefore, it is a general
consequence of any Fermi liquid description of quark-like
excitations.

The operator in (\ref{leading2}) is anti-Hermitian and leads to a
positive, semi-definite determinant since it anti-commutes with
$\gamma_5$. The corrections given in (\ref{treeL}) are all
Hermitian, so higher orders in the $1/ \mu$ expansion may
re-introduce complexity. The structure of the leading term plus
corrections is anti-Hermitian plus Hermitian, just as in the
original QCD Dirac Lagrangian with chemical potential. The leading
terms in the effective action for gluons (these terms are
generated when we match our effective theory, with energy cutoff
$\Lambda$, to QCD) also contribute only real, positive terms to
the partition function:
\begin{equation}
S_{\rm eff}(A)=\int{\rm
d}^4x_E\left({1\over4}F_{\mu\nu}^aF_{\mu\nu}^a +{M^2\over
16\pi}\sum_{\,\vec v_F}A_{\perp\mu}^{a}A_{\perp\mu}^{a}\right)
\ge0,
\end{equation}
where $A_{\perp}=A-A_{\parallel}$ and the Debye screening mass is
$M=\sqrt{N_f/(2\pi^2)}g_s\mu$\,.

Matching of hard gluon effects also leads to four-quark operators
in the effective theory. At asymptotic density, we can neglect
these operators, since forward scattering dominates Cooper pairing
interactions (due to Landau damping \cite{csc}). However, at lower
densities hard operators may be important. Matching effects due to
hard gluon exchange still lead to a positive action
for attractive channels, since they
arise from quasiparticle-gluon interactions which are originally
positive\footnote{A simple way to study the positivity of
four-quark operators is to replace them by a vector field with
trivial quadratic term $V_\mu^2$ which couples to quarks like the
original gluon: $V_\mu \bar{\psi} \gamma^\mu \psi$. Completing the
square, we see that the resulting path integral is positive
if the four-quark interactions are attractive.}.
Only interactions involving virtual anti-quarks lead to
non-positive interactions, and these are always suppressed by
powers of $\mu$.

Positivity of the measure allows for rigorous QCD inequalities at
asymptotic density. For example, inequalities among masses of
bound states can be obtained using bounds on bare quasiparticle
propagators. One subtlety that arises is that a quark mass term
does not lead to a quasiparticle gap (the mass term just shifts
the Fermi surface). Hence, for technical reasons the proof of
non-breaking of vector symmetries ~\cite{Vafa:1984xg} must be
modified. (Naive application of the Vafa-Witten theorem would
preclude the breaking of baryon number that is observed in the
color-flavor-locked (CFL) phase~\cite{Alford:1998mk}.) A
quasiparticle gap can be inserted by hand to regulate the bare
propagator, but it will explicitly violate baryon number. However,
following the logic of the Vafa-Witten proof, any symmetries which
are preserved by the regulator gap cannot be broken spontaneously.
One can, for example, still conclude that isospin symmetry is
never spontaneously broken. In the case of three flavors, one can
use the CFL gap as a regulator to show rigorously that none of the
symmetries of the CFL phase are broken at asymptotic density. On
the other hand, by applying anomaly matching conditions
\cite{anomaly}, we can prove that the axial symmetries {\it
are} broken. We therefore conclude that the CFL phase is the true
ground state for three light flavors at asymptotic density.

It may be possible to simulate dense QCD using the effective field
theory determinant
in place of the usual quark determinant. We know that this is a
good approximation at very high density, and it should remain a
good approximation as long as $\Lambda_{\rm QCD}$, the
characteristic scale of the dominant physics, is smaller than
$\mu$. At lower densities the weak coupling approximation no
longer holds, so (\ref{eliminate}) is no longer a good guide to the
higher order corrections, which are only constrained by symmetry
and projection requirements. However, we can use naive dimensional
analysis~\cite{Manohar:1983md} to estimate the coefficients of
the higher dimension operators.

An estimate of the size of corrections to the determinant from
higher orders can be obtained by considering the Euclidean
relation
\begin{equation}
\label{trln}
{\rm det M = e^{Tr ln M} = e^{ - \epsilon_0 V }}~~~,
\end{equation}
where $\epsilon_0$ is the vacuum energy density and V the volume
of the system. Corrections to the vacuum energy
can be estimated using conventional diagrammatic methods and naive
dimensional analysis. We find that corrections are
roughly suppressed by powers of $\frac{\alpha_s}{2 \pi}
\frac{\Lambda}{\mu}$.

To realize the effective theory (\ref{leading2}) directly on the
lattice, one can replace the plaquette $U_{n,n+\mu} \sim 1 + i a
A_{\mu} (n)$ by $U_+ \equiv e^{-iX} U e^{+iX}$ in the fermion
action, but not in the gauge action. In effect, one computes the
fermion determinant in the usual way, but as a function of $U_+$.
The momenta of the quasiparticle $\psi_+$ modes is simply the
residual momenta l, which is unrestricted except that its
magnitude must be small compared to $\mu$. One can impose this
condition on $\psi_+$ by choosing a lattice spacing $a_{\rm det}
>> \mu^{-1}$ in the quark determinant. However, a challenging
aspect of this determinant is that it is computed on the geometry
of a spherical shell rather than a ball: \begin{equation} d^3p ~=~
dp ~p^2 ~d\Omega ~=~ dl~(\mu+l)^2~ d\Omega ~\neq~ d^3l ~=~ dl
~l^2~ d\Omega~~~.
\end{equation}

To obtain the $e^{\pm iX}$ operators, one must first realize (and
presumably diagonalize) the velocity operator (\ref{velo}). Since
the momentum operator has a simple representation in coordinate
space, the most challenging aspect of $\vec{v}$ is the
normalization factor (the square root of the Laplacian in
(\ref{velo})), which is non-local. (A similar problem arises in
lattice models of chiral fermions.) One should probably
investigate a simpler method for directly enforcing the
normalization condition on $\vec{v}$. Note, however, that
$\vec{v}$ and $X$ are independent of the gauge field, so they need
be computed only once for each lattice.

Probably the most direct way to utilize our results on the lattice
is as follows. Imagine coupling dense quark matter to a background
gauge field A whose magnitude and derivatives are characterized by
a scale $\Lambda << \mu$. The leading part of the low-energy
effective theory (describing only Fermi surface modes) has a real
and positive determinant. The quark determinant (or equivalently,
its logarithm which is the effective action) can be expanded in
powers of $1/ \mu$, with the leading term real and positive. Since
the determinant is a functional of the gauge field and its
derivatives, the expansion will effectively be in powers of
$\lambda$ over $\mu$. This means that the ordinary lattice
determinant ${\rm det} [\gamma_{\mu}D_{\mu}+\mu\gamma_4]$ computed
in such backgrounds should be real and positive to leading order
in $1 / \mu$. Physically, the low-momentum gauge fields cannot
excite modes deep within the Fermi sea (such as anti-quarks) which
lead to complex contributions to the determinant. If we further
take $\Lambda >> \Lambda_{QCD}$, then the neglected effects of
higher momentum gluon modes are suppressed by powers of $\alpha_s
( \Lambda ) << 1$.

In order to restrict ourselves to gauge fields satisfying $\Lambda
<< \mu$ we need to couple the quark determinant living on a
lattice with spacing $a \sim 1/ \mu$ to a set of gauge fields
living on a coarser lattice of spacing $a' \sim 1/ \Lambda$.
Ordinary calculations of the dense matter quark determinant on
lattices with $a = a'$ will not yield a real, positive result
since the sub-leading terms in the $1/ \mu$ expansion of the
determinant are as large as the leading term.

\eject

\acknowledgments

We would like to thank M. Alford, T. Cohen, S. Hands, C. Kim, W.
Lee, T.-S. Park, K. Rajagopal, M. Schwetz and M. Stephanov for
useful discussions. The work of D.K.H. was supported in part by
the academic research fund of Ministry of Education, Republic of
Korea, Project No. BSRI-99-015-DI0114, and also by the KOSEF
through the Korea-USA Cooperative Science Program, 2000-111-04-2.
The work of S.H. was supported in part under DOE contract
DE-FG06-85ER40224 and by the NSF through through the USA-Korea
Cooperative Science Program, 9982164. D.K.H. and S.H. acknowledge
the hospitality of the APCTP Winter School on Dense Matter, where
this work was begun.


\begin{references}
\bibitem{ev}To demonstrate this, note that $\gamma_5$
anti-commutes with M, so if $M \phi = \lambda \phi$, then $M
\gamma_5 \phi = - \gamma_5 M \phi = - \lambda \gamma_5 \phi$.
\end{references}
\end{document}